\documentclass[prl,twocolumn,superscriptaddress,showpacs,floatfix,amsfonts]
{revtex4}
\usepackage{graphicx,graphics,color,epsfig}
\usepackage{bm}
\usepackage{amsmath}
\usepackage{amssymb}
\usepackage{epstopdf}
\begin{document}
\preprint{}

\title{The Bose-Fermi Kondo model with a singular dissipative spectrum:
Exact solutions and their implications}
\author{Jianhui Dai}
\affiliation{Zhejiang Institute of Modern Physics, Zhejiang University,
Hangzhou 310027, China}
\affiliation{Department of Physics and Astronomy, Rice University, Houston,
TX 77005, USA}
\author{Qimiao Si}
\affiliation{Department of Physics and Astronomy, Rice University, Houston,
TX 77005, USA}
\author{C. J. Bolech}
\affiliation{Department of Physics and Astronomy, Rice University, Houston,
TX 77005, USA}
\date{\today}
\begin{abstract}

Quantum dissipation induces a critical destruction of a Kondo
screened state, which is of interest in the contexts of quantum
critical heavy fermion metals and magnetic nanostructures.
The sub-ohmic Bose-Fermi Kondo model provides a setting to study
this effect. We find that this many-body problem is exactly
solvable when the spectrum of the dissipative bosonic bath,
$J(\omega)$, is singular, corresponding to $J(\tau)={\rm const.}$.
We determine the local spin correlation functions, showing 
that the singular {\it longitudinal} fluctuations of the
bosonic bath dominate over the transverse ones. Our results 
provide evidence that the local quantum critical solution, 
derived within the extended dynamical mean field approach 
to the Kondo lattice model, has a zero residual entropy.
\end{abstract}

\pacs{71.10.Hf,71.55.-i, 75.20.Hr,71.27.+a}

\maketitle

The Bose-Fermi Kondo model (BFKM)\cite{Smith,Sengupta} has become
of considerable interest in a number of different contexts,
including the extended dynamical mean field approach to the heavy
fermion quantum criticality\cite{Si1} and the dissipative effects
in mesoscopic structures\cite{Hur,Kirchner}. The model describes a
quantum impurity spin coupled simultaneously to a
conduction-electron bath\cite{Hewson} and to a dissipative bosonic
continuum\cite{Leggett,Sachdev}. The coupling
of the local moment
to conduction electrons leads to the Kondo screening effect.
The coupling to the bosonic bath, on the other hand,
causes the critical suppression of the Kondo effect.
The latter is the key ingredient
of the local quantum critical solution to the Kondo lattice
model\cite{Si1}, and 
has received some fairly direct
support from the experiments
in
heavy fermions\cite{Lohneysen,Gegenwart_Si_Steglich}.
In even broader contexts, models such as BFKM serve as prototype
systems to study
quantum dissipation and 
quantum coherence\cite{Leggett}.

The BFKM has a rich phase diagram, which is sensitive to
the form of 
the spectral density of the bosonic bath:
\begin{eqnarray}
J(\omega) \propto |\omega|^{1-\epsilon} ~\mathrm{sgn}(\omega)
~.
\label{dissipative_spectrum}
\end{eqnarray}
Most of the previous studies have focused on the sub-ohmic case
with $0<\epsilon<1$. The perturbative renormalization group (RG)
analysis\cite{Smith,Sengupta,Zhu,Zarand} shows that
spin-anisotropy is a relevant perturbation, and indeed the
Kondo-destroyed phase of the
Ising-anisotropic
BFKM\cite{Smith, Zhu, Grempel,Bulla,Glossop,Glossop_prb07}
differs 
qualitatively from that of 
the spin-isotropic
BFKM\cite{Sachdev00,Zhu.04}.
The cases of
singular bosonic spectrum [$J(\omega)$
being divergent in the zero-frequency
limit], $\epsilon>1$, have received much less attention.
(A singular bosonic spectrum might be realizable in some low
dimensional ferromagnetic single-electron
structures\cite{Kirchner}.) One may argue \cite{Si_04}
on general 
grounds that the large bosonic spectral weight at low energies 
implies that an infinitesimal coupling between the local moment
and the bosonic
bath will suppress the Kondo effect.
For the
Ising-anisotropic BFKM, the numerical renormalization group (NRG)
work of Glossop and Ingersent
\cite{Glossop_prb07}
has
indeed demonstrated this.

Several factors have motivated us to search for an exact solution
to BFKM. 
It will 
put the above perturbative, numerical, or saddle-point results on
a more firm ground, thereby elucidating the physics of 
not only the Kondo destruction but also the quantum dissipation/coherence
in general.
In addition, it may help resolve 
some apparent inconsistency concerning
the entropy of the local quantum critical solution 
to the Kondo lattice system.
Some general considerations suggest
a zero residual entropy (see below),
yet Kircan and Vojta \cite{Kircan.04} have interpreted
their results for a dynamical large-N limit of the spin-isotropic
BFKM as signaling a contrary conclusion. 

In this letter, we show that the BFKM with a singular bosonic bath,
corresponding to $\epsilon=2$, is exactly solvable.
The exact solution establishes that an infinitesimal coupling
between the local moment and the bosonic bath indeed causes
a Kondo destruction. 
Surprisingly, we find that the spin-isotropic problem 
has the same universal properties as the Ising-anisotropic one,
which implies that the longitudinal dissipative coupling plays
a dominant role.
This effect turns out to be 
missed by the large-N limit of
Ref.~\cite{Kircan.04} in an interesting way.

The 
Hamiltonian for 
the BFKM is
\begin{eqnarray}
H=H_{F}+H_{B}+H_{J}+H_{g}+H_h .
\label{hamiltonian_bfkm}
\end{eqnarray}
Here, $H_F=\sum_{k,\sigma}\varepsilon_{k}c^{\dagger}_ {k\sigma}
c_{k\sigma}$ and 
$H_B=\sum_{p,a}w_{p}\Phi^{\dagger}_{pa}\Phi_{pa}$
respectively describe a conduction-electron band
and a bosonic bath, and $\varepsilon_k$ and $w_{p}$ 
their corresponding dispersions. $H_J=(J_{\perp}/2) [s^-_c(0)S^+
+s^+_c(0)S^-]+J_{||}s^z_c(0)S^z$ 
specifies the Kondo coupling
between
a spin-$1/2$ local moment, $\vec{S}$,
and the conduction electron spin
$\vec{s}_c=\sum\frac{1}{2}c^{\dagger}_{k\sigma}\vec{\tau}_{\sigma,\sigma'}
c_{k\sigma'}$ (with $\vec{\tau}$ being the Pauli matrices)
and $H_g=\sum_{a=1}^3g_a\Phi^aS^a$ 
the impurity coupling to the bosonic field,
with $\Phi^a \equiv 
\sum_{p}(\Phi_{pa}+\Phi^{\dagger}_{-pa})$
at the impurity site $0$.
The dissipative bosonic spectrum,
$J(\omega)
\equiv \sum_{p}[\delta(\omega-w_{p})-\delta(\omega+w_{p})]$,
is
taken to have the form of Eq.~(\ref{dissipative_spectrum}). 
In order to probe the
local spin responses,
introduces a local magnetic field, $h$, 
via $H_h=hS^z$.

The partition function can be written in the path-integral
form.
The quantum impurity spin $\vec{S}$ is
represented by an $SU(2)$ coherent state $S\vec{\Omega}(\tau)$
with
the constraint $\vec{\Omega}^2(\tau)=1$. A Berry phase term
$\mathcal{S}_{B}[\vec{\Omega}]$ characterizes the quantum dynamics
of the impurity spin. Meanwhile, the bosonic bath can be traced out,
leading to a long-ranged interaction 
(along the $\tau$-dimension) for the impurity spin\cite{Smith,Sengupta}:
\begin{eqnarray}
Tr
\exp\{-\beta
(H_B+H_g)\}=Z_b
\int {\cal D}\vec{\Omega}
\exp\{-i
\mathcal{S}_{B}[\vec{\Omega}]\}~~~~\\
\times \exp\{\sum_{a=1}^3\int_0^{\beta}\int_0^{\beta}
d\tau_1d\tau_2\frac{g^2_aS\Omega^a(\tau_1)S\Omega^a(\tau_2)}
{4|\tau_2-\tau_1|^{2-\epsilon}}\},\nonumber
\end{eqnarray}
with $Z_b$ being the partition function of the bosonic bath. (Note
that the purely classical model defined along a chain,
$0<\tau<\beta$, is considered to be ill-defined for
$1<\epsilon \le 2$, since the long-ranged interaction makes 
the ground state energy per unit length diverge 
in the $\beta \rightarrow
\infty$ limit, but the quantum problem we consider is still
well-defined.\cite{note}) 
For $\epsilon=2$, the long-range interaction in the
last term becomes $\sum_{a=1}^3[\int_0^{\beta} d\tau
g_aS\Omega^a(\tau)/2]^2$, which can be further decomposed by
introducing a Hubbard-Stratanovich vector
$\vec{\lambda}=(\lambda_1,\lambda_2,\lambda_3)$:
\begin{eqnarray}
\exp\{\sum_{a=1}^3[\int_0^{\beta} d\tau
Sg_a\Omega^a(\tau)/2]^2\}=~~~~~~~~~~~~~~~~~~~\nonumber\\
\pi^{-3/2}\int d\vec{\lambda}\exp\{-\vec{\lambda}^2-\sum_{a=1}^3
g_a\lambda_a\int_0^{\beta}d\tau S\Omega^a(\tau)\}.
\end{eqnarray}
The total partition function of the BFKM, when rewritten in the
operator formalism, becomes
\begin{eqnarray}
Z=\mathrm{Tr} e^{-\beta H}&=&\pi^{-3/2}Z_b\int d\vec{\lambda}
e^{-\vec{\lambda}^2} \mathrm{Tr}
e^{-\beta \tilde{H}(\vec{\lambda})},\nonumber\\
\tilde{H}(\vec{\lambda})&=&H_F+ H_J+H_h+H_{\vec{\lambda}},
\label{Z_gaussian_averaging}
\end{eqnarray}
where $H_{\vec{\lambda}}=\sum_{a=1}^3g_a\lambda_aS^a$. Hence, we
are led to solve the BFKM, Eq.~(\ref{hamiltonian_bfkm}),
in terms of a pure Fermi-Kondo model in the presence of a
Gaussian-distributed magnetic field $\vec{\lambda}$,
along
with the external field $h$.

{\it The Bose-Kondo model}. We first consider 
the case with 
$g_a=g$ (for $a=1,2,3$)
and 
in the absence of conduction
electrons, {\it i.e.}, the 
$SU(2)$ Bose-Kondo model.
Here,
$\tilde{H}(\vec{\lambda})=hS^z+g\vec{\lambda}\cdot\vec{S}$. 
Using the parameterization 
$\vec{\lambda}=(\lambda\sin\varphi\cos\theta,
\lambda\sin\varphi\sin\theta,\lambda\cos\varphi)$,
we can specify the eigenvalues of $\tilde{H}(\vec{\lambda})$
as $\pm
g \tilde{\lambda}/2$, with
$\tilde{\lambda}=\lambda\sqrt{(1-y)^2+4y\cos^2\frac{\varphi}{2}}$
and $y=\frac{h}{g\lambda}$. Inserting these into
Eq.~(\ref{Z_gaussian_averaging}) leads to 
the impurity partition function,
$Z_{loc}= Z_b^{-1}Z$:
\begin{eqnarray}
Z_{loc}= \left (2\cosh\frac{h\beta}{2}+
\frac{g^2\beta}{2}\frac{\sinh\frac{h\beta}{2}}{h}
\right )e^{\frac{g^2\beta^2}{16}}.
\label{Z_su2}
\end{eqnarray}
It follows that the static 
local spin susceptibility is
\begin{eqnarray}
\chi_{loc}=\frac{1}{\beta}\frac{\partial^2 \ln Z_{loc}}{\partial
h^2}|_{h=0}
=\frac{\beta}{12}\frac{3+{g^2\beta^2}/{8}}{1+{g^2\beta^2}/{8}} 
~\,\stackrel{\beta \rightarrow \infty}{\longrightarrow}\,
~\frac{\beta}{12}
\label{chi_loc_su2_static}
\end{eqnarray}
The asymptotic behavior in the low-temperature limit has a Curie form,
$\chi_{loc} (T \rightarrow 0) = {\beta}/{12}$,
with a reduced Curie constant
($1/12$ instead of the free-spin value, $1/4$).

\begin{figure}[ht]
\epsfxsize=6.5cm \centerline{\epsffile{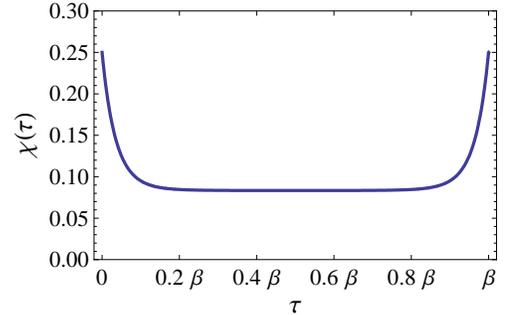}} \caption[]{The 
dynamical locl spin susceptibility of the $SU(2)$ Bose-only Kondo model
at $\epsilon=2$
(
with $g=1$ and $\beta=10$).}
\label{fig:dysus}
\end{figure}

The dynamical local spin-spin correlation function,
$\chi_{loc}(\tau) \equiv \langle S^z(\tau)S^z(0)\rangle_{loc}$,
is
obtained from
a Gaussian averaging:
$\chi_{loc}(\tau) =\left ( \pi^{-3/2} / Z_{loc} \right )
\int d\vec{\lambda}e^{-\vec{\lambda}^2}
A(\lambda)$,
where 
$ A(\lambda)
=Tr e^{-\beta
\tilde{H}[\vec{\lambda}]}S^z(\tau)S^z(0)$.
The trace amounts to 
$\sum_{n,m}e^{-E_n\beta}e^{(E_n-E_m)\tau}|\langle
n|S^z|m\rangle|^2$, where $n$ and $m$ run over
all the eigenstates of $\tilde{H}[\vec{\lambda}]$:
$|+\rangle=
(\cos\frac{\varphi}{2}e^{-i\frac{\theta}{2}},
\sin\frac{\varphi}{2}e^{i\frac{\theta}{2}})$, $|-\rangle=
(-\sin\frac{\varphi}{2}e^{-i\frac{\theta}{2}},
\cos\frac{\varphi}{2}e^{i\frac{\theta}{2}})$.
It follows that
$
A(\lambda)
=\frac{1}{2}[\cos^2\varphi\cosh\frac{g\lambda\beta}{2}+
\sin^2\varphi \cosh\frac{g\lambda(\beta-2\tau)}{2}]$
and,
in turn,
\begin{eqnarray}
\chi_{loc}(\tau)
=\frac{1}{12}[1+
\frac{2+\frac{g^2(\beta-2\tau)^2}{4}}{1+\frac{g^2\beta^2}{8}}
e^{-\frac{g^2}{4}\tau(\beta-\tau)} ]
~\mathop{\longrightarrow}_{\beta \rightarrow \infty}^{\tau \rightarrow
\beta/2} ~\frac{1}{12} .
\label{chi_loc_su2}
\end{eqnarray}
In the asymptotic low-temperature and long-time limit
($\tau\rightarrow \beta/2$, $\beta \rightarrow \infty$),
it also has a Curie form.
The full result
is also illustrated in
Fig.~\ref{fig:dysus}.

We next consider the 
Ising case,
{\it i.e.},
$g_1=g_2=0, g_3=g$. 
The impurity partition function is now
\begin{eqnarray}
Z_{loc}=2\cosh\frac{h\beta}{2}e^{\frac{g^2\beta^2}{16}} .
\label{Z_Ising}
\end{eqnarray}
Both the static and dynamical local spin susceptibility 
have simple Curie forms:
$\chi_{loc} ={\beta}/{4}$,
and $\chi_{loc}(\tau)={1}/{4}$.

To summarize, the local spin response
of both
the $SU(2)$ and the Ising cases
has a Curie form in the low temperature limit.
For the $SU(2)$ problem, the fact that it has the same
universal behavior 
as the Ising case implies that it too is controlled by the 
$g^*=\infty$ fixed point; in other words, 
the longitudinal fluctuations
dominate over the transverse ones.
This is in contrast to the 
$\epsilon < 1$ case, where 
the $SU(2)$ and Ising problems are controlled by 
a finite-$g^*$ fixed
point and a $g^*=\infty$ one, respectively.
\cite{Smith,Sengupta,Zhu,Zarand}
Our exact result for the $\epsilon =2$ case reveals the link between 
a singular bosonic spectrum and dominant longitudinal fluctuations.
We therefore expect that,
for the $1<\epsilon
< 2$ cases too,
the $SU(2)$ problem will be controlled by the
fixed point at $g^*=\infty$.

{\it The Bose-Fermi Kondo model}.
We now 
incorporate the Kondo coupling as well. The spin-isotropic case
requires the usage of Bethe-ansatz method, which will be discussed
elsewhere\cite{Dai}. Here we consider 
the 
case in which the pure fermionic Kondo part is placed at 
its Toulouse point\cite{Hewson} and, moreover,
the impurity-boson coupling
is purely Ising. 
Since the spin-anisotropy in the impurity-bosonic
coupling is
irrelevant and our results will demonstrate that an infinitesimal
coupling to the bosonic bath leads to the Kondo destruction, the
spin-isotropic BFKM should behave similarly.

When the Kondo couplings take the Toulouse values, the model
Eq.~(\ref{Z_gaussian_averaging})
can be mapped\cite{Hewson} to the following non-interacting spinless
resonant level model (RLM):
$H_T(\lambda)=\sum_{\vec{k}} \left [ 
\varepsilon_{\vec{k}}c^{\dagger}_{\vec{k}}
c_{\vec{k}}+V(c^{\dagger}_{\vec{k}}d+h.c.) \right ]
+\varepsilon_d
(d^{\dagger}d-\frac{1}{2})$.\nonumber\\
Here, the impurity spin is represented by $S^z\rightarrow
d^{\dagger}d-1/2$, 
$S^+ \rightarrow d^{\dagger}$,
and 
$S^- \rightarrow d$,
with $d$ describing 
a spinless fermion.
The hybridization
$V$
is proportional to $J_{\perp}$, while
$\varepsilon_d=g\lambda+h$.  After integrating out the
fermionic bath, we obtain $ Z_{T}=\pi^{-1/2}\int d\lambda
e^{-\lambda^2}\mathrm{Tr} e^{-\beta H_T(\lambda)} =Z_cZ_{loc}$.
Here, $Z_{c}$ is the partition function of the electron bath,
and $Z_{loc}=\pi^{-1/2}\int d\lambda e^{-\lambda^2}Z_{loc}[\lambda]$
with
\begin{eqnarray}
Z_{loc}[\lambda]=&& 2 
\exp\{-\beta\int_{-D}^{D}
\frac{d\omega}{\pi}n(\omega)
\arctan\frac{\Gamma}{\omega-\epsilon_d}\} 
\nonumber\\
&&\cdot \cosh(\frac{\beta\epsilon_d}{2})  ,
\label{Z-RLM}
\end{eqnarray}
where $n(\omega)=[1+\exp(\beta \omega)]^{-1}$ is the Fermi
function, $\Gamma=\pi\rho_0 V^2$ the bare resonance width,
and $\rho_0$ the conduction-electron density of states
at the Fermi energy. Without loss of generality, we assume a flat
band for the electrons, taking the usual limit of a large
bandwidth $D$ ($\gg \Gamma$).

Eq.~(\ref{Z-RLM}) has been derived with care,
containing not only
the usual phase shift contribution \cite{Hewson},
but also a residual
atomic term. Indeed, 
it 
recovers the results
for both the Ising bosonic Kondo model ($\Gamma=0$)
and the conventional RLM ($g=0$).
When both couplings $\Gamma$ and $g$ are non-zero, the Gaussian
averaging over $\lambda$ complicates the problem.
We focus on
the zero-field static local susceptibility.
It is 
straightforward to show that
\begin{eqnarray}
\chi_{loc}=\frac{\int_{-\infty}^{\infty}d\lambda
e^{-\lambda^2}\chi_{loc}[\lambda] Z_{loc}[\lambda]}
 {\int_{-\infty}^{\infty}d\lambda e^{-\lambda^2}Z_{loc}[\lambda]}|_{h=0},
 \label{chi0}
\end{eqnarray}
where $\chi_{loc}[\lambda]=\{\frac{\partial}{\partial
h}M_{loc}[\lambda]+\beta M_{loc}^2[\lambda]\}$ and
$M_{loc}[\lambda]=\frac{1}{\beta}\frac{\partial \ln
Z_{loc}[\lambda]}{\partial h}$. At low temperatures, 
we use the asymptotic approximation $n(\omega)\approx
\Theta(-\omega)$ (Sommerfeld expansion, valid up to
corrections of order of $T^2/\Gamma$) and integrate over
$\omega$, obtaining $\chi_{loc}[\lambda]=\frac{1}{\pi\Gamma}\frac{\Gamma^2}
{\Gamma^2+g^2\lambda^2}+\frac{\beta}{\pi^2}\arctan^2\frac{g\lambda}
{\Gamma}$ and $Z_{loc}[\lambda]=\exp\{\frac{\beta
g\lambda}{\pi}\arctan\frac{g\lambda}{\Gamma}
-\frac{\beta\Gamma}{2\pi}\ln[1+(\frac{g\lambda}{\Gamma})^2]\} Z'$,
where $Z'$ is independent of $\lambda$ in the limit
$D/\Gamma\rightarrow\infty$.

\begin{figure}[ht]
\epsfxsize=8.0cm \centerline{\epsffile{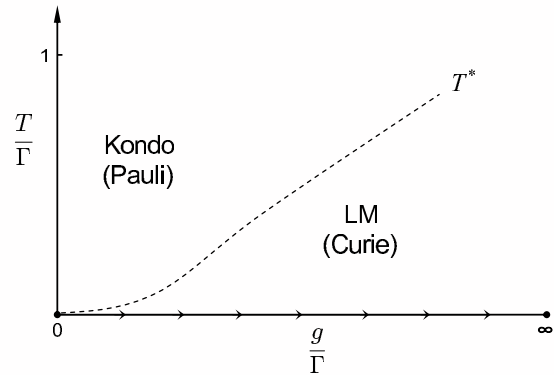}} 
\caption[]{The
phase diagram of the 
present BFKM.
$\chi_{loc}$ has the Curie or Pauli forms when $T\ll T^*$ or
$T^*\ll T\ll\Gamma$;
for $g \ll \Gamma$,
$T^*\approx g^2/(2\pi\Gamma)$.
}
\label{fig:phase}
\end{figure}

In the weak boson-impurity coupling regime, $g/\Gamma \ll 1$, the
$\lambda$-integrals in Eq.(\ref{chi0}) can be asymptotically
approximated by
introducing a cut-off $\Lambda \approx \Gamma/g \gg 1$. Within the
cut-off ($|\lambda| \ll \Lambda$, called region I below),
the dummy variable
$\frac{g\lambda}{\Gamma}$ is always small such that expansions
like $\arctan\frac{g\lambda}{\Gamma}\approx
\frac{g\lambda}{\Gamma}$ are appropriate. Outside the cut-off
($|\lambda| \gg \Lambda$, region II), $\frac{g\lambda}{\Gamma}$ is large
enough and one has $\arctan\frac{g\lambda}{\Gamma}\approx
\mathrm{sgn} (\lambda)\frac{\pi}{2}$. 
$\chi_{loc}[\lambda]$ then
becomes 
$\frac{1}{\pi\Gamma}+\frac{\beta}{\pi^2} \left (
\frac{g\lambda}{\Gamma} \right )^2 $
and $\frac{\beta}{4}$ for $|\lambda| \ll \Lambda$ and
$|\lambda| \gg \Lambda$, respectively.
$Z_{loc}[\lambda]$ contains a dimensionless combination,
$\alpha=\frac{\beta g^2}{2\pi\Gamma}$.
A characteristic temperature scale, $T^*=\frac{
g^2}{2\pi\Gamma}$ (corresponding to $\alpha=1$),
arises, separating two limiting
temperature regimes with distinct asymptotic behavior 
for $\chi_{loc}$.
In the low temperature regime, {\it i.e.}, $T
\ll
T^*
$ ( or $\alpha\gg 1$), 
the $\lambda$-integration is dominated by the region-II contribution,
and $\chi_{loc}\approx\frac{\beta}{4}$. 
In the higher temperature regime,
$T^* \ll T \ll \Gamma$ (or $\alpha\rightarrow 0$),
on the
other hand, 
the region-I contribution dominates,
and 
$\chi_{loc}\approx\frac{1}{\pi\Gamma}$. A
smooth, but broad, crossover takes place 
around $T \sim T^*$.
The phase diagram is illustrated in
Fig.~\ref{fig:phase}.

In short, for a fixed non-zero Kondo coupling (and, hence,
non-zero $\Gamma$), the local spin susceptibility in the
low-temperature limit turns to 
a Curie form 
for any non-zero $g$.
This implies that an infinitesimal $g$ causes a
destruction of the Kondo effect;
the BFKM is
controlled by the fixed point associated with the Bose-only Kondo
model solved in the previous section. 

{\it Implications for the EDMFT solution of the Kondo lattice
model}.
The extended dynamical mean field 
approach to the Kondo lattice model
is studied through a
BFKM with self-consistent spectra for the bosonic and
fermionic baths. In the local quantum critical
solution,\cite{Si1,Grempel,Glossop07,Zhu07} the
self-consistent
bosonic bath has a spectrum with $\epsilon=1^{-}$.

For $\epsilon<1$, we have the standard sub-ohmic situation in
which an unstable fixed point (critical, referred to as ``C'')
separates the Kondo and LM
fixed points.\cite{Zhu,Zarand}
NRG studies of the Ising-anisotropic
BFKM\cite{Glossop_prb07}
found that, for $1<\epsilon<2$, an infinitesimal $g$
leads to Kondo destruction.
(Similar results were inferred from an O(N)
representation of the Ising-anisotropic BFKM\cite{Si_04}.)
In other words, at $\epsilon$ exactly equals to $1$, ``C'' and
the Kondo fixed points are merged and the impurity entropy will
vanish. We expect the impurity entropy to be a smooth function
of $\epsilon$ (which is consistent with the NRG result\cite{Bulla-private}),
so the impurity entropy should vanish at $\epsilon=1^-$ as well.
It follows that the corresponding solution to the Kondo lattice 
has a vanishing residual entropy, which has been experimentally
seen in YbRh$_{\rm 2}$Si$_{\rm 2}$, a prototype quantum critical
heavy fermion metal.\cite{Custers}.

The situation is very different in the large-N limit discussed in
Ref.~\cite{Kircan.04}, where ``C'' and the Kondo fixed points are
still separated even for $1<\epsilon<2$.\cite{Zhu_04_note} 
Moreover, the impurity
entropy, at both the ``C'' and the LM fixed points, are finite for
the entire range of $0<\epsilon<2$, including at $\epsilon=1$.

Our exact results imply that the large-N limit becomes problematic
when the bosonic bath is singular.
The
local susceptibilities of both the LM and ``C''
fixed points of the large-N limit were shown to have
$\chi_{loc}(\tau)
\sim 1/\tau^{\eta}$, with $\eta=\epsilon$ for the
entire range of $0<\epsilon\leq 2$ \cite{Kircan.04};
the structure of fixed points remains the same 
over this range $0<\epsilon\leq 2$ \cite{Kircan.04}.
This would
give rise to $\eta=2$ for $\epsilon=2$. But our exact solution
for the $SU(2)$ case [{\it cf.}, Eq.~(\ref{chi_loc_su2}) and
Fig.~\ref{fig:dysus}]
shows that $\eta=0$ for $\epsilon=2$. The problem lies in
the large-N-limit's under-estimation of the longitudinal
fluctuations, which is of order $1/N$. 
As we discussed earlier,
when the bosonic
bath has a singular spectrum ($\epsilon>1$) and for any
finite-N, the strong longitudinal fluctuations dominates
over the transverse fluctuations, 
and the spin-isotropic
problem 
fall in the same universality class as the Ising case.
This implies that the structure of the fixed points of the
spin-isotropic BFKM changes as $\epsilon$ passes
through $1$.

Keeping track of the longitudinal
fluctuations is also important to enforce the 
Griffiths' bound\cite{Griffiths},
which requires $\chi_{loc}(\tau)$ {\it not} to decay faster than
the interaction, $1/\tau^{2-\epsilon}$.
This bound is violated (for $\epsilon > 1$)
by the $N=\infty$ result,\cite{Kircan.04}
$\chi_{loc}(\tau) \sim 1/\tau^{\epsilon}$. 
A contribution of order $1/\tau^{2-\epsilon}$ 
to $\chi_{loc}(\tau)$ occurs at order $1/N$,
which is missed in the $N=\infty$ limit. On the other hand,
our exact result for the $SU(2)$ problem at $\epsilon=2$, 
$\chi_{loc}(\tau) \sim 1/\tau^{0}$,
satisfies the
Griffiths' bound.

In conclusion, we have presented the first exact solution to the
Bose-Fermi Kondo model. Our results display the physics of
dissipation-induced Kondo destruction, an effect that is of
extensive current interest in the understanding of quantum
criticality in heavy fermion metals. Our results also establish
the important role that the longitudinal fluctuations of the
dissipative bath plays in the
Kondo destruction, which has been missed in an
interesting way in a large-N formulation of this problem. The
new understandings achieved here have allowed us to elucidate
the issue of zero residual
entropy of the local quantum critical solution to the Kondo
lattice models.

We thank R. Bulla, M. T. Glossop and, particularly,
M. Vojta, for useful discussions,
and the NSF Grant No. DMR-0706625
and 
the Robert A. Welch Foundation
for support. One of us (J.D.)
has also been supported by the NSF of China and the
PCSIRT(IRT-0754) of the Chinese Ministry of Education.


\begin{thebibliography}{99}
\bibitem{note}
Note however
the overall pre-factor $e^{g^2\beta^2/16}$ in
the partition functions, Eqs.~(\ref{Z_su2},\ref{Z_Ising}).
Its origin lies in the renormalization of $Z_b$ by $g$.
Some regularization procedure may be introduced
to cancel this pre-factor.
This, however, does not affect the local spin responses.
\end{thebibliography}
\end{document}